%Paper: hep-th/9311178
%From: abdalla@surya11.cern.ch (Elcio Abdalla)
%Date: Tue, 30 Nov 93 09:30:07 +0100
%Date (revised): Sun, 5 Dec 93 12:29:13 +0100

%%%%%%%%%%%%%%%%%%%%%%%%%%%%%%%%%%%%%%%%%%%%%%%%%%%%%%%%%%%%%%%%%%%%%%%%%%%%%%%
\magnification 1200
\font\eightrm=cmr8
\font\eighti=cmmi8
\font\eightsy=cmsy8
\font\eightbf=cmbx8
\font\eighttt=cmtt8
\font\eightit=cmti8
\font\eightsl=cmsl8
\font\sixrm=cmr6
\font\sixi=cmmi6
\font\sixsy=cmsy6
\font\sixbf=cmbx6
\catcode`@11
\newskip\ttglue

\def\eightpoint{\def\rm{\fam0\eightrm}
\textfont0=\eightrm \scriptfont0=\sixrm \scriptscriptfont0=\fiverm
\textfont1=\eighti \scriptfont1=\sixi \scriptscriptfont1=\fivei
\textfont2=\eightsy \scriptfont2=\sixsy \scriptscriptfont2=\fivesy
\textfont3=\tenex \scriptfont3=\tenex \scriptscriptfont3=\tenex
\textfont\itfam=\eightit \def\it{\fam\itfam\eightit}
\textfont\slfam=\eightsl \def\sl{\fam\slfam\eightsl}
\textfont\ttfam=\eighttt \def\tt{\fam\ttfam\eighttt}
\textfont\bffam=\eightbf
\scriptfont\bffam=\sixbf
\scriptscriptfont\bffam=\fivebf \def\bf{\fam\bffam\eightbf}
\tt \ttglue=.5em plus.25em minus.15em
\normalbaselineskip=6pt
\setbox\strutbox=\hbox{\vrule height7pt width0pt depth2pt}
\let\sc=\sixrm \let\big=\eightbig \normalbaselines\rm}
\newinsert\footins
\def\newfoot#1{\let\@sf\empty
  \ifhmode\edef\@sf{\spacefactor\the\spacefactor}\fi
  #1\@sf\vfootnote{#1}}
\def\vfootnote#1{\insert\footins\bgroup\eightpoint
  \interlinepenalty\interfootnotelinepenalty
  \splittopskip\ht\strutbox % top baseline for broken footnotes
  \splitmaxdepth\dp\strutbox \floatingpenalty\@MM
  \leftskip\z@skip \rightskip\z@skip
  \textindent{#1}\footstrut\futurelet\next\fo@t}
\def\fo@t{\ifcat\bgroup\noexpand\next \let\next\f@@t
  \else\let\next\f@t\fi \next}
\def\f@@t{\bgroup\aftergroup\@foot\let\next}
\def\f@t#1{#1\@foot}
\def\@foot{\strut\egroup}
\def\footstrut{\vbox to\splittopskip{}}
\skip\footins=\bigskipamount % space added when footnote is present
\count\footins=1000 % footnote magnification factor (1 to 1)
\dimen\footins=8in % maximum footnotes per page

\def\ref#1{$^{#1}$}
\def\flex{\raise 6pt\hbox{$\leftrightarrow $}\! \! \! \! \! \! }
\def\oversome#1{ \raise 8pt\hbox{$\scriptscriptstyle #1$}\! \! \! \! \! \! }

\newbox\bigstrutbox
\setbox\bigstrutbox=\hbox{\vrule height10pt depth5pt width0pt}
\def\bigstrut{\relax\ifmmode\copy\bigstrutbox\else\unhcopy\bigstrutbox\fi}
\def\refer[#1/#2]{ \item{#1} {{#2}} } 
\def\rev<#1/#2/#3/#4>{{\it #1\/} {\bf#2}, {#3}({#4})}
\def\boxit#1{\vbox{\hrule\hbox{\vrule\kern3pt
\vbox{\kern3pt#1\kern3pt}\kern3pt\vrule}\hrule}}

\def\2figure#1#2#3#4{\vbox{ \hrule width#1truecm \hbox{\vrule height#2truecm
\hskip #1truecm   
\vrule height#2truecm }\hrule width#1truecm \hbox{\vrule\vbox{\hsize #1truecm 
\baselineskip=10pt 
\noindent\strut#3}\vrule}\hrule width#1truecm 
\hbox{\vrule\vbox{\hsize #1truecm 
\baselineskip=10pt 
\noindent\strut#4}\vrule}\hrule width#1truecm  }}
\def\3figure#1#2#3#4#5{\vbox{ \hrule width#1truecm \hbox{\vrule height#2truecm
\hskip #1truecm   
\vrule height#2truecm }\hrule width#1truecm \hbox{\vrule\vbox{\hsize #1truecm 
\baselineskip=10pt 
\noindent\strut#3}\vrule}\hrule width#1truecm 
 \hbox{\vrule\vbox{\hsize #1truecm 
\baselineskip=10pt 
\noindent\strut#4}\vrule}
\hrule width#1truecm \hbox{\vrule\vbox{\hsize #1truecm 
\baselineskip=10pt 
\noindent\strut#5}\vrule}\hrule width#1truecm  }}

\def\sqr#1#2{{\vcenter{\hrule height.#2pt
   \hbox{\vrule width.#2pt height#1pt \kern#1pt
    \vrule width.#2pt}
    \hrule height.#2pt}}}

% Here are my additional definitions:

\def\smin{\,\raise 0.06em \hbox{${\scriptstyle \in}$}\,}
\def\smsubset{\,\raise 0.06em \hbox{${\scriptstyle \subset}$}\,}

\def\Natural{\hbox{\hskip 1.5pt\hbox to 0pt{\hskip -2pt I\hss}N}}

\def\Rational{\hbox{\hbox to 0pt{\hskip 2.7pt \vrule height 6.5pt
                                  depth -0.2pt width 0.8pt \hss}Q}}
\def\Real{\hbox{\hskip 1.5pt\hbox to 0pt{\hskip -2pt I\hss}R}}
\def\Complex{\hbox{\hbox to 0pt{\hskip 2.7pt \vrule height 6.5pt
                                  depth -0.2pt width 0.8pt \hss}C}}
%%%%%%%%%%%%%%%%%
% definitions for the second book
%tcap1

\def \E {{{\rm e}}}

%%%%%%%%tcap2

%%%%%%%%tcap3
\def \1ok{{1\over \kappa ^2} }

\def \3dslim {{\rm DS}\!\!\!\!\!\!\!\!\lim }
\def \4dslim {{\rm DS}\!\!\!\!\!\!\!\!\!\!\lim }

\def \2kk{\left( \matrix {2k\cr k\cr }\right) }
\def \Rs4{{R^k\over 4^k} }

%%%%%%%%%tcap4
\def \1ok{{1\over \kappa ^2} }

%%%%%%%%%%tcap5

%%%%%%%%%%tcap6
%\def \DSLim {{\rm DS}\!\!\lim }
%%%%%%%%%%tcap7
%nada
%%%%%%%%%%tcap8
%nada
%%%%%%%%%appb

%%%%%%%%%%%%%%%%%%%%%%%%%%%%%%%%%%%%%%%%%%%%%%%%%%%%%%%%%%%%%%%%%%%%%%%%%%%%%%%
\input epsf.tex
\def \E{{\rm e}}
\def\flex{\raise 6pt\hbox{$\leftrightarrow $}\! \! \! \! \! }
\nopagenumbers
\centerline{Exact S matrices and extended supersymmetry}
\vskip 1.5cm
\centerline {\bf E. Abdalla\newfoot {${}^*$}{Work supported by
FAPESP. Permanent address: Instituto de F\'\i sica - USP, C.P. 20516, S.Paulo, 
Brazil.}, M.C.B. Abdalla\newfoot {${}^\dagger $}{Supported by CNPq. 
Permanent address: Instituto de F\'\i sica Te\'orica - UNESP, R. Pamplona 145,
CEP-01405, S.Paulo, Brazil.}}
\centerline { CERN-Theory Division}
\centerline { 1211 Geneve 23, Switzerland.}
\vskip 1cm
\centerline {\bf Abstract}
\vskip.4cm
We study the constraint imposed by supersymmetry on the exact $S$-matrix of 
$\Complex P^{n-1}$ model, and compute a non-trivial phase factor in the 
relation between the $S$-matrix and one of the supersymmetry generators.
We discuss several features connected with the physical interpretation of
the result. The supersymmetry current is studied in such context as well,
and we find some operators appearing in the conservation equation of the 
supersymmetry current.
The relation with the literature on the subject is also discussed.
\vfill
\eject
\countdef\pageno=0 \pageno=1
\newtoks\footline \footline={\hss\tenrm\folio\hss}
\def\folio{\ifnum\pageno<0 \romannumeral-\pageno \else\number\pageno \fi}
\def\advancepageno{\ifnum\pageno<0 \global\advance\pageno by -1
\else\global\advance\pageno by 1 \fi}
\noindent 
The issue of supersymmetry has been playing a central role in the development of
quantum field theory since two decades. Supersymmetric grand unified theories
are the most promissing candidates to extend the standard  model of strong and
eletro-weak interations. In the string domain supersymmetry is essential in
order to define a tachyon free theory. Nevertheless, it must be broken already
at intermediate energies, in order to take into account the observed particle
spectrum as revealed from experiments.

Integrable theories on the other hand offer an extraordinary theoretical 
laboratory to test ideas in quantum field theory. They provide examples
of models which are exactly soluble at the level of $S$-matrix. Thus the 
analysis of supersymmetric integrable theories is worthwhile to pursue
in order to analyse ideas.

On the other hand, there has been recently a large number of results concerning
integrable $N=2$ supersymmetric models. Gepner's conjecture\ref{1}, linking 
these models to perturbations of superconformally invariant theories with
Landau-Ginzburg potentials have been put forward by a number of
authors\ref{2,3}. For the time being, this program has been given quite a solid 
basis, relying on results from singularity (catastrophe) theory\ref{4,5} on the 
one hand, in order to obtain a classification of the Landau-Ginzburg theories, 
and computations based on the thermodynamic Bethe ansatz obtaining\ref{6} the 
central charge associated with the exact $S$-matrices of the given $N=2$ 
supersymmetric theory, on the other hand.

The afore mentioned relations lead to a complete classification of the
integrable $N=2$ supersymmetric theories in terms of perturbed Landau-Ginzburg
potentials\ref{7}.

In this spirit, a number of examples have been analysed. The supersymmetric
generalizations of coset space models - or Kazama Suzuki models\ref{8}, have 
been studied. These coset models are characterized in terms of a cohomology 
for the elements of the chiral primary rings\ref{7}, and some models are
indeed represented as Landau-Ginzburg theories. Fendley and Intriligator\ref{6}
considered minimal theories perturbed by the least relevant operator. They
considered the peculiar phenomenon of fractional fermion number in 
two-dimensions, where in polymer systems a fermion can be distorted into $n$
solitons, which carry, as a consequence, fermion number $1/n$. This sort of 
phenomenon plays a central role in our discussion. In fact, they stated that 
the fractional fermion number is crucial for obtaining the correct soliton 
content and $S$-matrix. 

The $S$-matrix corresponding to a $Z(n)$ supersymmetric theory has also been
computed in [3]. A previous calculation of the same $S$-matrix was given by
[9], and both results are different. The latter did not contain supersymmetry
as a constraint for the $S$-matrix.

We do not discuss whether any of these results is either correct or incorrect,
since from our point of view, a dynamical principle is required to choose which
of them will describe the given physics. Indeed, a $Z(n)$ invariant model 
displaying the $S$-matrix given by [9] has to display a breakdown of 
supersymmetry, at least if it is to be realized locally. On the other hand, the
results from [3] are supersymmetric, and must describe  models which are
supersymmetric at the quantum level. 

Here we consider a similar problem for the computation of the exact $S$-matrix
of the supersymmetric $\Complex P^{n-1}$ model. In this case, we verify that 
the $S$-matrix  fullfills the supersymmetry constraint, only if one considers
carefully certain phase factors associated with the statistics of the
asymptotic fields.  We first 
make several consistency checks of the proposed $S$-matrix, as well as the $1/n$
perturbative check. We consider in detail the case of $\Complex P^1$ and 
 compare it with the $O(3)$ $S$-matrix, and we see that the equivalence of 
$\Complex P^1$ and $O(3)$ models, verified in the purely bosonic case at the
$S$-matrix level\ref{10}, as well as by means of numerical simulations\ref{11} 
seems to break down in the supersymmetric case. We shall discuss the 
consequences and possible causes of such behavior, as well as whether one
can argue in favor of a  supersymmetry anomaly in the $\Complex P^{n-1}$ models.

We have computed some candidates to a supersymmetry anomaly in the context of
the $1/n$ expansion of the model. Whether these candidates are anomalies or
not can only be decided by the study of the cohomology of such 
operators\ref{12}. Further explanations of the results obtained
may be a failure in the description of asymptotic states, or a non standard
action of the supersymmetry generators.

The supersymmetric $\Complex P^{n-1}$ model is defined by the Lagrangian
density\ref{13}
$$
{\cal L}= \overline {D_\mu z} D^\mu z +\overline \psi 
iD\! \llap /\psi  +{f\over 2n} \left( \overline \psi \gamma _\mu \psi
\right) ^2 +{f \over 2n} \left[ \left( \overline \psi \psi \right) ^2
-\left( \overline \psi \gamma _5\psi \right) ^2\right] \quad , \eqno(1)
$$
with  the constraints $\overline z_i z_i ={n\over 2f}$ and $\overline z_i 
\psi _i = 0 = \overline\psi _i z_i$, where $D_\mu = \partial _\mu - {2f\over n}
\overline z\partial _\mu z$. 

This Lagrangian is invariant under the supersymmetry transformation
$$
\eqalignno{
\delta z_i & = \overline \epsilon \psi_i \quad, \cr
\delta \psi_i & = - i \not \!\! D z_i \epsilon + {1\over 2}z_i\left[ \overline
\psi \psi + \overline \psi \gamma_5 \psi \gamma_5  + \overline \psi \gamma^\mu
\psi \gamma_\mu \right] \epsilon \quad ,&(2)\cr}
$$
where $\epsilon $ is a Grassmmann valued parameter. This symmetry will be the 
central issue of the present discussion. The $1/n$ expansion of this model is 
well known and will not be discussed here. Further details will be given in a 
more detailed publication.

We consider the exact $S$-matrix of the supersymmetric $\Complex P^{n-1}$ 
model. We make the ansatz (the backward particle-antiparticle scattering 
vanishes as a consequence of the conservation  of the non-local charges; we 
have already used this fact in the ansatz\ref{14})
$$
\eqalignno{ 
\langle z_\beta (\theta '_1)z_\delta (\theta '_2)\vert
z_\alpha (\theta _1) z_\gamma (\theta _2)\rangle = &
[v_1(\theta )\delta _{\alpha \beta }\delta _{\gamma \delta } + 
v_2(\theta )\delta _{\alpha \delta }\delta _{\gamma \beta }]
\delta (\theta _1-\theta '_1)\delta (\theta _2-\theta '_2) \cr
& +[v_1(\theta )\delta _{\alpha \delta }\delta _{\gamma \beta }+
v_2(\theta )\delta _{\alpha \beta }\delta _{\gamma \delta }]
\delta (\theta _1-\theta '_2)\delta (\theta _2-\theta '_1)
\; ,\cr
\langle \psi_\beta (\theta '_1)\psi_\delta(\theta'_2)\vert 
\psi_\alpha(\theta _1)\psi_\gamma (\theta _2)\rangle =&
[u_1(\theta )\delta _{\alpha \beta }\delta _{\gamma \delta } + 
u_2(\theta )\delta _{\alpha \delta }\delta _{\gamma \beta }]
\delta (\theta _1-\theta '_1)\delta (\theta _2-\theta '_2) \cr
& -[u_1(\theta )\delta _{\alpha \delta }\delta _{\gamma \beta }+
u_2(\theta )\delta _{\alpha \beta }\delta _{\gamma \delta }]
\delta (\theta _1-\theta '_2)\delta (\theta _2-\theta '_1)\; ,\cr
\langle \psi_\beta (\theta '_1)z_\delta (\theta '_2)\vert 
\psi_\alpha (\theta _1)z_\gamma (\theta _2)\rangle  = &
[c_1(\theta )\delta _{\alpha \beta }\delta _{\gamma \delta } + 
c_2(\theta )\delta _{\alpha \delta }\delta _{\gamma \beta }]
\delta (\theta _1-\theta '_1)\delta (\theta _2-\theta '_2) \cr
& +[d_1(\theta )\delta _{\alpha \delta }\delta _{\gamma \beta }+
d_2(\theta )\delta _{\alpha \beta }\delta _{\gamma \delta }]
\delta (\theta _1-\theta '_2)\delta (\theta _2-\theta '_1)\; ,&(3)\cr 
\langle z_\beta (\theta '_1)\overline z_\delta (\theta'_2)\vert z_\alpha 
(\theta _1) \overline z_\gamma (\theta _2)\rangle  = &[v_1(i\pi\! -\! \theta )
\delta _{_{\alpha \beta }}\delta _{_{\gamma \delta }} + v_2(i\pi\! -\!\theta )
\delta _{_{\alpha \gamma }}\delta _{_{\delta \beta }}]
\delta (\theta _1 \!-\!\theta '_1)\delta (\theta _2\!-\!\theta '_2) \; ,\cr
\langle \psi_\beta (\theta '_1)\overline \psi_\delta (\theta '_2)\vert
\psi_\alpha (\theta _1) \overline \psi_\gamma (\theta _2)\rangle = 
&[u_1(i\pi\! -\!\theta )\delta _{_{\alpha \beta }}\delta _{_{\gamma \delta }} + 
u_2(i\pi \!-\!\theta )\delta _{_{\alpha \gamma }}\delta _{_{\delta \beta }}]
\delta (\theta _1\!-\!\theta '_1)\delta (\theta _2\!-\!\theta '_2)\; ,\cr
\langle z_\beta (\theta '_1)\overline z_\delta (\theta '_2)\vert \psi_\alpha
(\theta _1) \overline \psi_\gamma (\theta _2)\rangle = &
[d_1(i\pi\! -\!\theta )\delta _{_{\alpha \beta }}\delta _{_{\gamma \delta }} + 
d_2(i\pi \!-\!\theta )\delta _{_{\alpha \gamma }}\delta _{_{\delta \beta }}]
\delta (\theta _1\!-\!\theta '_1)\delta (\theta _2\!-\!\theta '_2)\; ,\cr
\langle z_\beta (\theta '_1)\overline \psi_\delta (\theta '_2)\vert z_\alpha
(\theta _1) \overline \psi_\gamma (\theta _2)\rangle = &
 [c_1(i\pi\! -\!\theta )\delta _{_{\alpha \beta }}\delta _{_{\gamma \delta }} + 
c_2(i\pi \!-\!\theta )\delta _{_{\alpha \gamma }}\delta _{_{\delta \beta }}]
\delta (\theta _1\!-\!\theta '_1)\delta (\theta _2\!-\!\theta '_2)\; .\cr}
$$ 

Proceeding as usually, one defines the action of the quantum non-local charge on
asymptotic states, and one arrives at the solution 
$$
v_1(\theta )={\sin  \Bigl( \!{\theta \over 2i}\!-\!{\pi \over n}\!\Bigr)
\over \sin  {\theta \over 2i}}c_1(\theta )\; ,\; 
u_1(\theta ) = {\sin  \Bigl(\! {\theta \over 2i}\!+\!{\pi \over n}\!\Bigr)
\over \sin  {\theta \over 2i}}c_1(\theta )\; ,\;
d_1(\theta ) =-{\sin  \Bigl(\! {\pi \over n}\!\Bigr)\over \sin  
{\theta \over 2i}}c_1(\theta )\; ,\; \eqno(4)
$$
and
$$
\eqalignno{
v_2(\theta )&={2i\pi \over n\theta }v_1(\theta )\quad ,\quad
u_2(\theta )={2i\pi \over n\theta }u_1(\theta )\; ,\cr
c_2(\theta )&=-{2i\pi \over n\theta }c_1(\theta )\quad ,\quad 
d_2(\theta )=-{2i\pi \over n\theta }d_1(\theta )\; ,\;&(5)\cr }
$$
which contains also the bound state structure of the model as we will comment
later. Equations (5) are a consequence of the non-local
conservation laws, while (4) are a consequence of
factorization of the amplitude for the scattering of the two
bosons and one fermion, as well as crossing. Finally, we 
determine $c_1(\theta )$ from unitarity.
$$
c_1(\theta )=\prod _{l=0}^\infty {\Gamma \Bigl( {\theta \over
2\pi i}+{1\over n}+l\Bigr) \Gamma \Bigl( 1-{\theta \over 2\pi
i}+l\Bigr) \Gamma \Bigl( {\theta \over 2\pi i}-{1\over
n}+l\Bigr) \Gamma \Bigl( 1-{\theta \over 2\pi i}+l\Bigr)  \over
\Gamma \Bigl( 1-{\theta \over 2\pi i}+{1\over n}+l\Bigr) \Gamma
\Bigl( {\theta \over 2\pi i}+l\Bigr) \Gamma \Bigl( 1- {\theta
\over 2\pi i}-{1\over n}+l\Bigr) \Gamma \Bigl( {\theta \over 2\pi i}+l\Bigr) }
\quad .\eqno(6)
$$
This shows the  existence of  bound states, with the spectrum given by the
fusion rule by  
$$
m_l=m{\sin  {l\pi \over n}\over \sin  {\pi \over n}}\quad ,\quad
l=1,\cdots ,n-1\quad .\eqno(7)
$$
In this theory, a bound state of $n-1$ fermions is equivalent to
a antiboson, i.e.
$$
\overline z_\delta ={1\over (n-1)!}\epsilon _{\delta \alpha
_1\cdots \alpha _{n-1}}\psi_{\alpha _1}\cdots \psi_{\alpha
_{n-1}}\quad ,\eqno(8)
$$
while the bound state of $n-2$ fermions and a boson is a antifermion 
$$
\overline \psi_\alpha  ={1\over (n-1)!}\epsilon _{\alpha  \beta 
_1\cdots \beta _{n-1}}\psi_{\beta _1}\cdots \psi_{\beta
_{n-2}}z_{\beta _{n-1}}\quad .\eqno(9)
$$
From these formulae it follows that the particle-antiparticle backward
scattering amplitudes vanish identically. Using the $1/n$ expansion we can 
check the leading terms in the above amplitudes (see [14] for details).

Let us consider the ansatz (3) for the $S$-matrix. The action of the 
supersymmetry charge on asymptotic states is easy to obtain. We shall only  
consider the action of the supersymmetry charge in one sector, in such a way 
that complications arising from topological sectors\ref{15} are immediately 
dismissed. See also [3] and [6]. We shall come back to this point later. 
The action of the charge on  a two particle state is given by
$$
\eqalign{
Q\vert z_i(\theta_1)z_j(\theta_2)\rangle &= \E^{{\theta\over 4}} \vert
\psi_i(\theta_1)z_j(\theta_2)\rangle + \E^{i\epsilon}\E^{-{\theta\over 4}} \vert
z_i(\theta_1)\psi_j(\theta_2)\rangle \quad ,\cr
Q\vert \psi_i(\theta_1)z_j(\theta_2)\rangle &= \E^{{\theta\over 4}} \vert
z_i(\theta_1)z_j(\theta_2)\rangle + \E^{i\epsilon'}\E^{-{\theta\over 4}} \vert
\psi_i(\theta_1)\psi_j(\theta_2)\rangle \quad ,\cr}\eqno(10)
$$
where $\epsilon$ and $\epsilon'$ are related with the fermionic content of the 
particles $z_i$ and $\psi_j$. Usually one expects $\epsilon =0$ and 
$\epsilon'=\pi$.

We need also the $S$-matrix elements
$$
\eqalign{
S\vert z_i(\theta_1)z_j(\theta_2)\rangle =& \left( v_1(\theta)\delta_{ik}
\delta_{jl} + v_2(\theta) \delta_{il}\delta_{jk}\right) \vert z_l(\theta_2) z_k
(\theta_1)\rangle \quad ,\cr  
S\vert \psi_i(\theta_1)z_j(\theta_2)\rangle =& \left( c_1(\theta)\delta_{ik}
\delta_{jl} + c_2(\theta) \delta_{il}\delta_{jk} \right)\vert z_l(\theta_2) 
\psi_k (\theta_1)\rangle + \cr  
&+ \left( d_1(\theta)\delta_{il}\delta_{jk} + d_2(\theta) \delta_{ik}
\delta_{jl} \right)\vert \psi_k(\theta_2) z_l (\theta_1)\rangle \quad .\cr}
\eqno(11)
$$
 
We analyse the commutator of the $S$-matrix with the supersymmetry charge $Q$, 
that is, we impose the condition
$$
\langle z_k(\theta'_1) \psi_l(\theta'_2) \vert QS \vert z_i(\theta_1) z_j
(\theta_2)\rangle = \langle z_k(\theta'_1) \psi_l(\theta'_2) \vert S Q 
\vert z_i(\theta_1) z_j (\theta_2)\rangle \quad .\eqno(12)
$$
The above equation leads to  very simple constraints on the $S$-matrix elements
(11). We are led to the equations
$$
\eqalign{
 \E^{i\epsilon}\E^{{\theta\over 4}}v_1(\theta) &=  \E^{{\theta\over 
4}}c_1(\theta) +
 \E^{i\epsilon}\E^{-{\theta\over 4}}d_1(\theta) \quad ,\cr
\E^{{\theta\over 4}}v_2(\theta) &= \E^{{\theta\over 4}}c_2(\theta) +
 \E^{i\epsilon}\E^{-{\theta\over 4}}d_1(\theta)\quad , \cr
\E^{-{\theta\over 4}}v_1(\theta) &= 
\E^{{\theta\over 4}}d_1(\theta) +
\E^{i\epsilon}\E^{-{\theta\over 4}}c_1(\theta) \quad ,\cr
 \E^{-{\theta\over 4}}v_2(\theta) &= 
\E^{{\theta\over 4}}d_2(\theta) +
\E^{i\epsilon}\E^{-{\theta\over 4}}c_2(\theta) \quad .\cr} \eqno(13)
$$

The question which arises is whether such equations are compatible with the
solution given in eqs. (3-5). If we consider the relation
$$ \E^{i\epsilon} \E^{\theta\over 2} { \E^{i\epsilon}v_1(\theta) - c_1(\theta) 
\over d_1(\theta) } = -  \E^{i\epsilon}
\E^{\theta\over 2}{ \E^{i\epsilon}\sin\left({\theta\over 2i} - {\pi\over 
n}\right) - \sin {\theta\over 2i} \over \sin{\pi\over n}}\quad , \eqno(14)
$$
we immediately see that it is not one as required by the first equation in
(14) unless $\epsilon ={i\pi\over n}$! On the other hand, we will see that the 
candidate to a 
supersymmetry anomaly is proportional to $\phi_5 \gamma_5 c$.
Witten\ref{13} has already proven long ago, that $\phi_5$ differs by multiples 
of $2\pi/n$ between ``in" and ``out" states. Therefore, the anomaly found in 
order $1/n$ can be intepreted as a result of the above equation relating the 
``in" and ``out" states.

Also connected with this  question is  the fact that, as shown in 
[14], the $N=2$ supersymmetry algebra contains central terms due to the soliton
content of the theory. Notice however, that we have only used the $N=1$
subalgebra. In such a case, we can extract, out of the full algebra
$$
\eqalignno{
\{ Q'_\alpha, Q'_\beta \} & = \{ \overline Q'_\alpha, \overline Q'_\beta \} =0
\quad , &(15)\cr
\{ Q'_\alpha, \overline Q'_\beta \} & = \gamma^\mu _{\alpha\beta}P_\mu  +
\delta_{\alpha\beta }\phi + \gamma_{5\alpha\beta}\phi_5\quad , &(16)\cr}
$$
the subalgebra
$$
Q_+= Q'_2 + {Q'}^\dagger_2= Q'_2 + (\overline Q \gamma^0)_2\quad ,\eqno(17)
$$
such that $ Q^2_+ = P_+$; indeed, we have only considered $Q_+$. The question
 about the $U(1)$ charge carried by the supersymmetry charge has to be addressed
 at this point. In fact, from the physical interpretation of the model, and
especially from the bound state structure (7-9), one learns that the
elementary fields carry a fractional extra charge equal to $\pm {1\over n}$.
Therefore the fractional value $\epsilon =\pi/n$ has been found. However, we
found satisfactory explanation for the phase difference between ``in" and
``out" states.

When fermions are coupled to the model, the first problem that appears is the
fact that the gauge field has now a short distance behaviour. The fermions
eliminate the long range force by means of a screening mechanism. If the
fermions are minimally coupled, all fermionic degrees of freedom  decouple. On 
the other hand, the gauge field acquires a mass. This means that a dynamical 
Higgs mechanism took place. The bosonic particle is no longer confined. The 
$S$-matrix of the minimally coupled $\Complex P^{n-1}$ model confirms that no 
bound state exists in this case. For the supersymmetric model a similar 
feature is realized. However, due to the large
number of fermionic degrees of freedom, some fermion fields survive the Higgs
mechanism. In the case of minimally coupled model, there is no bound state, as
one verifies from the absence of poles in the physical region, looking at the
solution for the $S$-matrix describing bosonic scattering, and verified by
means of the $1/n$ expansion. The surviving fermionic degrees of freedom are
massive, and  show a different behavior under chiral transformations as compared
with the fermions one started with. 

Therefore, in the supersymmetric model, one verifies that the interaction of the
gauge field with the fermions reveals a mechanism similar to that involving the
interaction of the chiral Gross Neveu model with a gauge field, which 
assimilates the Goldstone boson of the model, becoming heavy.

In the case of the pure Gross Neveu model, consideration of the physical
degrees of freedom leads to a set of $SU(n)$ massive interpolating fields with
spin ${1\over 2}\left(1-{1\over n}\right)$, while the chirality carrying fields
decouples. This permits an identification of the antifermion as a bound state of
fermions. For the $\Complex P^{n-1}$ model the combination of the Higgs 
mechanism for the
gauge field with the non-trivial $\Complex P^{n-1}$ interactions, leads now to a
generalization of the mechanism according to which antiparticles are bound 
states of particles (see (7-9)). As in the interaction of gauge fields with the
chiral Gross Neveu fermions, the chirality carrying part of the fermions is
absorbed into the gauge field, leaving room for the bound state condition. In
the sigma model interaction of $\Complex P^{n-1}$ model with fermions, the
gauge invariant field $\epsilon ^{\mu\nu}F_{\mu\nu}$ as well as the chiral
density $\overline \psi \gamma_5 \psi$ get a non-vanishing expectation value in
the presence of a $\theta$-term added to the Lagrangian. These first
expectation values however vanish in the supersymmetric limit, while the second
is maximal, signalizing the screening of the external field. This means that
the chirality carrying field spurionizes on the gauge invariant subspace, and
can not be seen in the physical spectrum.

The first example  that some unusual fact is on the way, concerns the example
of the $\Complex P^1$ model. The purely bosonic model is equivalent to the
$O(3)$  non-linear $\sigma$ model. This fact is well established. At the level 
of $S$-matrix it is realized by the fact that the $O(3)$ elementary particle 
is a particle-antiparticle bound state. The bound state pole at the $\Complex 
P^1$ level lies at the edge of the physical region, a fact that can be traced 
back to the confining properties of the model. All these facts have been 
checked by means of numerical computations\ref{11}.

In the supersymmetric $\Complex P^{1}$ model, the solution of the constraint 
$\overline z_i \psi_i=0$ can be written as $\overline \psi_i = \epsilon _{ij}
z_j \chi$, where $\chi$ is a anticommuting field. However, from the solution 
(4-6) one verifies that $\chi$ is empty, and can at most be a Klein factor 
assuring anticommutativity of the fermion field. This is easily verified at 
the $S$-matrix level by means of the identity
$$
c_1^{(n=2)}(i\pi -\theta) = {i\pi - \theta\over \theta} {\cos {\theta\over 2i}
\over \sin {\theta\over 2i}} \, c_1^{(n=2)}(\theta)\quad .\eqno(18)
$$ 
With such an identification, it is difficult to reassure supersymmetry! But the
worst fact concerns the expected equivalence between $\Complex P^1$ and $O(3)$ 
models, which should be realized by the identifications
$$
\varphi_i = \sigma _i^{\alpha\beta} z_\alpha\overline z_\beta\quad ,\quad 
{\rm and} 
\quad \chi_i  = \sigma_i ^{\alpha\beta}(z_\alpha \psi^+_\beta + z^+_\alpha
\psi_\beta)\quad .\eqno(19)
$$

Absence of fermion-boson backward scattering for the susy $O(3)$ model, is
required from the supersymmetric constraints. However it implies that the
$d_1(\theta)$ amplitude must vanish. 

Let us  consider the classical conservation of the supersymmetric
current, and latter quantize each step, in order to see, at the end, at which
point a different result with respect to the classical case appears. We have
$$
J_\mu = \overline {D_\nu z_i} \gamma^\nu \gamma_\mu \psi_i \quad , \eqno(20)
$$
whose divergence is given by
$$
\eqalignno{
\partial ^\mu J_\mu & = \overline {D^2z_i}\psi_i - {1\over 2}\epsilon ^{\mu\nu}
F^{\mu\nu}\overline z_i \psi_i+\overline{\not \!\! D z_i}\not \!\! D \psi_i\cr
&= \!\left(\! {\alpha\over \sqrt n}\! -\! m^2\right)\overline z_i \psi_i + \! 
{1\over \sqrt n}\left( \overline \psi_i c \right)\psi_i  - \overline {\not 
\!\! D z_i}\! \left( \! m \! +\! {\varphi\over \sqrt n} \!+ \!i \gamma_5 
\varphi_5\right)\psi_i \!+ \!{i\over \sqrt n} \overline {\not \!\! D z_i }z_i 
c \, ,&(21)\cr}
$$
where we used the equations of motion.

The divergence of the supersymmetric current vanishes upon use of the following
facts; first one uses a Fierz transformation of the second term. Later one uses
the (classical) definitions
$$
\sqrt n \,m + {\varphi}  = {f\over \sqrt n} \overline \psi \psi \quad ,\quad
\varphi_5 = i{f\over \sqrt n}\overline \psi \gamma_5 \psi \quad , \quad
c = i{2f\over \sqrt n} \overline z {\not \!\! D }\psi \quad ,\eqno(22)
$$
and finally, the constraint $\overline z _i \psi _i = 0 $. We shall verify 
that the last equation fails under a very special condition,
spoiling conservation of the supersymmetry current.

From this point on we shall use the quantization of the field operators as 
defined by the  BPHZ scheme\ref{16,17}. Thus, composite field operators
are made finite by means of subtractions around zero external momenta.

Identifications of the type (22) are rather subtle in the framework of the 
$1/n$ expansion, and are related to the fact that for a special vertex 
$N[\overline z\psi ]$, the diagram  obtained by joining the $z$ and
$\psi$-lines into a $\overline c \overline z \psi$ vertex, which is the origin
of a $c$-line ending at another vertex of the same type, is cancelled by the 
diagram where this second vertex originats directly from the $N[\overline z
\psi]$ vertex. However this is not true when this vertex is oversubtracted.

There are special terms which require special care, arising from the mass term 
in the equations of motion. Indeed, these terms in principle do not require
renormalization, but in order to achieve cancellation of the terms resulting
from the equations of motion the renormalization  of such terms presents
problems.

Indeed, in terms of subtraction of diagrams around zero momenta, one defines
various quantum composite operators $N_\delta[{\cal O}] $ corresponding to a
classical operator ${\cal O}$. The index $\delta$ defines how a diagram
containing the special vertex $N_\delta[{\cal O}]$ has to be
subtracted\ref{16,17}. For a diagram with $n_\psi$ external fundamental 
fermions, $n_z$ external $z$-bosons, $n_c\,$ $c$-lines, etc. we have, for the 
superficial degree of divergence of a generic diagram $(\gamma)$ the 
expression [16,17,18]
$$
\delta(\gamma) = 2 - {1\over 2}n_\psi - 2n_z - 2 n_\alpha - {3\over 2} n_c -
\cdots + \sum _a (\delta_a -2)\quad .\eqno(23)
$$

Moreover the quantum operator $ N_\delta[{\cal O}] $ satisfies
$$
\partial _\mu N_\delta[{\cal O}] = N_{\delta +1} [\partial _\mu {\cal O}]\quad
.\eqno(24)
$$

A minimally subtracted finite quantum operator has  $\delta$ equal to the
naive dimension of the operator ${\cal O}$. This is the least value of $\delta$
such that diagrams containing ${\cal O}$ be finite. We consider thus the normal
product (or quantum operator) $J_\mu = N_{3/2}[\overline {\not \!\! D z}
\gamma_\mu \psi]$. Its divergence is given by terms of dimension 5/2, that is
$$
\eqalignno{
\partial ^\mu J_\mu & =   N_{5/2}[ \partial ^\mu ( \overline {\not \!\! D z}
\gamma_\mu \psi) ]\quad , \cr
&=  N_{5/2}[\overline { D^2 z}\psi] + N_{5/2}[\epsilon_{\mu\nu} F^{\mu\nu}
\gamma_5 \overline z \psi] + N_{5/2}[\overline {\not \!\! D z} \not \!\! D \psi]
\quad . &(25)\cr}
$$

The last step is to insert the equation of motion inside the normal product. In
$1/n$ perturbation this is a standard graphical procedure [17]. Each time one 
has a factor of the inverse propagator, i.e. $i\!\!\not \!\! \partial -m$
for fermions, or $ \partial ^2 + m^2$ for bosons a propagator line is deleted. 
The resulting diagram is obtained replacing the inverse propagator by the field
in the next vertex at which one would arrive following the (deleted) line. In
fact, one obtains the terms appearing in the classical equation of motion. 

The final step has already been outlined previously and as we mencioned, the
step involving the constraint $ \overline z_i \psi_i =0 $ is dangerous. Indeed,
suppose that we have an over-subtraction of this operator, that is, we consider 
the operator $N_{1/2+\delta'} [\overline z_i \psi_i]$ where $\delta'$ is a
non-negative number.

For the operator we discussed previously we have, from a diagram with one
external $c$-line the contribution
$$
N_{5/2}[\overline z_i \psi_i]_{1c} ={i\over 4\pi m} c - {1\over 8\pi m^2} 
\not \!\partial c \quad .\eqno(26)
$$

Since the classical operator $\overline z \psi$ is invariant under
supersymmetry transformations (2), the above equation signals a
non-supersymmetric quantization of this operator.

In fact, such quantization is mandatory. Indeed,  to obtain the
quantum equations of motion in the way above described, we have to add the mass
term, to obtain the inverse propagator, thus we add and subtract to (25) the
terms
$$
N_{5/2}[m^2\overline z\psi]+ i N_{5/2}[m \overline {\not \!\! D z } \psi]\quad
.\eqno(27)
$$

The classical value of the first term, that is $ N_{1/2} [m^2\overline z\psi] $
is zero, while the second term, $ N_{3/2}[m\overline {\not \!\! D z} \psi]$
is just $mc$, and together with the term $\varphi c$ will cancel the term
$\overline \psi \psi$ in the equation of motion of the fermion. Therefore we
have to compute the extra subtractions of the terms (27), which are well known 
in the framework of BPHZ renormalization, corresponding to the so-called 
Zimmermann identities:
$$
N_{{\rm dim }({\cal O} +\delta)}[{\cal O}] = N_{{\rm dim }{\cal O}}[{\cal O}] +
\sum _i r_i N_{{\rm dim }({\cal O} +\delta)}[{\cal O}_i]\quad ,\eqno(28)
$$
where ${\cal O}_i$ are all possible terms with dimension ${\rm dim }( {\cal O}
+\delta)$. The easiest way to compute these terms in the right hand side is to
explicitely separate the last (extra) subtractions of the left hand side for
each specific diagram. In this way we shall compute $r_i$, in a $1/n$
perturbativ expansion. We have to compute the extra subtractions.

For the terms at lowest order in $1/n$ we arrive at the partial result
$$
\partial ^\mu J_\mu = {1\over 4\pi }\left[ m + {1\over 2}\not\! \partial+{1
\over \sqrt n} (\varphi+i\gamma_5 \varphi_5)\right] c + \cdots \quad ,\eqno(29)
$$
where the dots are coming from the diagrams to be considered later.

From the diagram with two external $\psi$-lines and one external
$c$-line  in the figures\newfoot{\ref\dagger}{A thin line represents a 
fermion, a thick one a boson, a dotted-traced line the $c\overline c$ 
propagator, and the wavy line one of the propagators of the remaining 
auxiliary fields. See [18], or [13] for details.} below.  For the first
diagram we have: 
$$
{\cal C}_{\alpha\beta,\gamma\delta}^{\Gamma \Gamma'} = m \! \int \! \! 
{{\rm d}^2 k\over (2\pi )^2} {i\over k^2 - m^2} \!\left[ \! (\! \not k - m )
{i\over \not k - m} \Gamma \right]_{\alpha \beta}\!\!{1\over F(k) + i/\pi} 
{\cal O}_{\Gamma\Gamma'}\!\!\left( \! \Gamma' {\not k + m\over k^2 - m^2}
\right)_{\gamma\delta} , \eqno(30)
$$
\centerline{\epsfysize=2.5cm\epsfbox{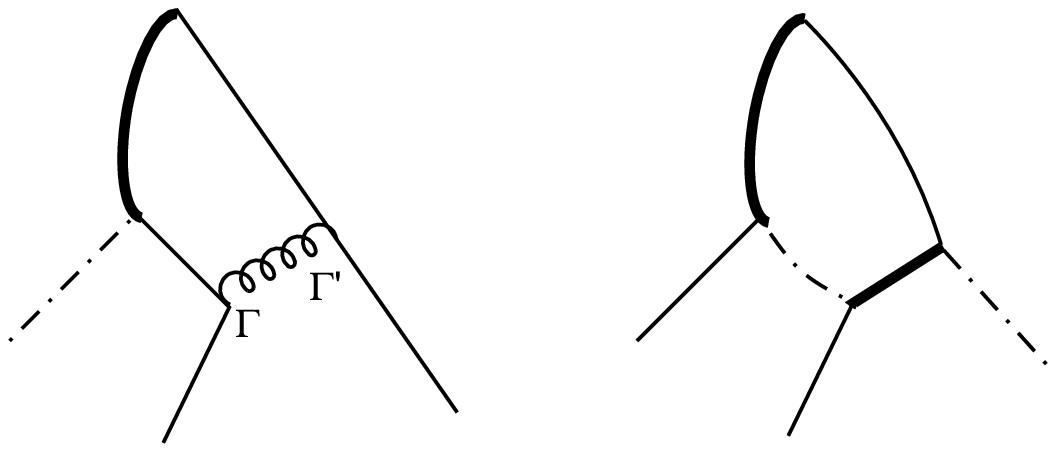}}
\nobreak
\centerline{Diagrams contributing to the equation (29), with two external 
fermions.}

\noindent which will contribute with a term of the type $\sum _{\Gamma\Gamma'} 
{\cal C} _{\alpha\beta,\gamma\delta} \psi_\beta \overline
\psi_\gamma i c_\delta$, where the factor $ {\cal O}_{\Gamma\Gamma'}$ depends 
on which of the propagators $(\Delta_\varphi, \Delta_{5}, \Delta_{\mu\nu}, 
\Delta_\mu ^5)$ we are considering. We define the constant
$$
{\cal W} = m^2 \int {{\rm d}^2k\over (2\pi)^2}{1\over \left[ F(k) + {i\over \pi}
\right] (k^2-m^2)^2}\quad .\eqno(31)
$$

We have the following contributions

a) $\Gamma= \Gamma' =i \quad,\quad{\cal O}_{\Gamma\Gamma'}=-1\quad,\quad
\Longrightarrow$ ${\cal C}_1 = -{\cal W}\left(\overline \psi c\right)\psi $

b) $\Gamma= \Gamma' =-\gamma_5 \quad,\quad{\cal O}_{\Gamma\Gamma'}=+1\quad,
\quad \Longrightarrow$ ${\cal C}_2 = -{\cal W}\gamma_5\psi\overline
\psi\gamma_5 c$
 
c) $\Gamma= -\gamma_5\, ,\, \Gamma' =i\gamma_\mu \quad,\quad 
{\cal O}_{\Gamma\Gamma'}= {2mi\over k^2}\epsilon _{\mu\nu} k^\nu\quad,
\quad \Longrightarrow$ ${\cal C}_3 = + 2 {\cal W} \gamma_5 \psi \overline \psi
\gamma_5 c $

d) $\Gamma= i\gamma_\mu \, ,\, \Gamma' =-\gamma_5 \quad,\quad 
{\cal O}_{\Gamma\Gamma'}= -{2mi\over k^2}\epsilon _{\mu\nu} k^\nu\quad,
\quad \Longrightarrow$ ${\cal C}_4 = {\cal W} \gamma^\mu \psi \overline \psi
\gamma_\mu c $

e) $\Gamma= i\gamma_\mu \, ,\, \Gamma' =i\gamma_\nu \quad,\quad 
{\cal O}_{\Gamma\Gamma'}= -\left( g_{\mu\nu} - {k_\mu k_\nu\over k^2}\right)
\quad,\quad \Longrightarrow$ ${\cal C}_5 = -{1\over 2} {\cal W} \gamma^\mu \psi 
\overline \psi\gamma_\mu c \, $.

Adding all these contributions and performing a Fierz transformation, we have
$$
\sum {\cal C} = {\cal W} \left[ {1\over 2}\overline \psi \psi - {1\over 2}
\overline \psi \gamma_5 \psi \gamma_5  + \overline \psi \gamma^\mu \psi
\gamma_\mu \right] c \quad .\eqno(32)
$$

Consider now the second diagram. It is given by
$$
{\cal E} _{\alpha\beta,\gamma\delta}= i^2 \int {{\rm d}^2k\over (2\pi)^2}
\left( {2(\not k - 2m) \over F(k) + i/\pi} \right)_{\alpha \beta} \left(
{i\over k^2-m^2} \right)^2 \left( {i\over \not k -
m}\right)_{\gamma\delta}\quad ,\eqno(33)
$$
which will contribute as 
$$
{\cal E} = {\cal E} _{\alpha\beta,\gamma\delta} \overline \psi_\alpha 
\psi_\beta i c_\delta \quad . \eqno(34)
$$

Defining now
$$
{\cal W}' = m^2 \int {{\rm d}^2k\over (2\pi)^2}
 {1 \over \left[ F(k) + i/\pi \right] (k^2 - m^2)^3} \quad  \eqno(35)
$$
we find
$$
{\cal E} = {\cal W}' \left[ 4\overline\psi\psi -\overline \psi \gamma^\mu \psi
\gamma_\mu \right] c  \quad . \eqno(36)
$$
Therefore we have
$$
\left[ ({1\over 2} {\cal W} + 4{\cal W}' ) \overline \psi \psi + ({\cal W} 
+{\cal W}')
\overline \psi \gamma_\mu \psi \gamma^\mu - {1\over 2}{\cal W} \overline \psi
\gamma_5 \psi \gamma_5 \right] c \quad . \eqno(37)
$$

This equation can be used to choose the coupling constant in $ \sqrt n \, m +\!
\varphi = {f\over \sqrt n}\overline \psi \psi$ in order to cancel the $\varphi$
contribution to the right hand side of equation (29), but not the $\varphi_5$ 
and $\overline \psi \gamma_\mu \psi$ contributions at the same time. Therefore 
we arrive at
$$
\partial ^\mu J_\mu = {\cal W}_1 \overline \psi \gamma_5 \psi  \gamma_5 c + 
{\cal W}_2 \overline \psi \gamma_\mu \psi \gamma^\mu c \quad , \eqno(38) 
$$
which we claim to be a candidate to the supersymmetry anomaly. The detailed 
computation of ${\cal W}_1$ and ${\cal W}_2$ depends on the normalization of 
the current.

In fact, one expects that one of the supersymmetry currents be conserved, since
the model can be formulated in an explicitely supersymmetric fashion\ref{19}.
                      
A further indication that the supersymmetric $\Complex P^{n-1}$ model has a
higher symmetry preventing the boson fermion back scattering amplitude to be
vanishing when the constraint $ \overline z \psi=0 $ is strongly realized,
comes from the perturbative computation. Indeed, up to fourth order in a 
perturbative computation, there is no contribution for $d_1(\theta)$ in (3).
                                                                  
We have verified that the $S$-matrix of the $\Complex P^{n-1}$ model is
supersymmetric if one considers non trivial phase factors for asymptotic
states. This may be a consequence of a possible supersymmetry anomaly 
which would appears due to the quantization of the constraint $\overline z_i 
\psi_i=0$. Nevertheless, this can also be understood from 
further properties of the model. Indeed, the implicit Higgs mechanism shifting 
the pole of a gauge field away from the origin, implies a set of bound states 
connecting tightly particles and antiparticles, and more importantly, fermions 
and bosons. Therefore the statistics got lost. The Klein factor used in 
two-dimensions to reassure the correct statistics after bosonization does not 
seem to be enough for a rebuilding of the supersymmetry algebra.

Although the bound state structure might be seen at first look as a
consequence of supersymmetry, generalizing the structure of the chiral Gross 
Neveu model, a concrete realization of supersymmetry at the level of the 
$S$-matrix fails if statistics is not taken into account. The supersymmetry 
algebra, if reconstructed,  should act on the fields with non trivial phase
factors due to asymptotic statistics.

Consequences for higher dimensional physics are vast. In three
dimensions\ref{19} there are two phases in the model, and in one of them the
issue of supersymmetry displays a behavior similar to the one presented 
here\ref{20}. This opens the possibility of describing particles
at very high energy in terms of supersymmetric grand unified theories, while at
sufficiently low energies supersymmetry breaks as a consequence of dynamics.

The fact that composite operators might be responsable for supersymmetry 
breaking has been conjectured before\ref{21}. In fact some candidates have been 
written for such composite (anomalous) operators. Although we did not check in
detail the relation to this line of works, it is conceivable that both issues 
can be related.

Several consistency checks of the results have been performed and will be 
reported elsewhere. First the result is non-perturbative. We have checked that
$d_1$ is zero up to fourth order in the coupling constant, while being non
zero in the $1/n$ expansion. Moreover the break of supersymmetry is minimal, in
the sense that at lowest order in $1/n$ the commutator of the supersymmetry 
charge with the $S$ operators is a constant, which is presumably necessary
in order to avoid further problems. In the $\Complex P^1$ case the $d$ term in 
the $S$-matrix may imply non equivalence with the supersymmetric $O(3)$ model. 
Moreover, the algebra underlining the bound state structure does not seem to
accomodate supersymmetry transformations in a simple way.

Finally, we should also point out that consistency of the supersymmetry
transformations, $U(1)$ current conservation and soliton type boundary 
condition for the auxiliary fields as initially pointed out by Witten\ref{13},
is consistent with our results. A supersymmetry transformation of the
naive $U(1)$ current, $J_\mu = i\overline z\partial_\mu z +
\overline\psi\gamma_\mu\psi$ leads to an expression of the type
$$-{1\over 2\sqrt n}\overline c\gamma_\mu\epsilon
-{1\over 2\sqrt n}\overline \epsilon\gamma_\mu c -{i\over 2}\overline J_\mu
^{susy}\epsilon +{i\over 2}\overline\epsilon J_\mu^{susy}$$

We expected this expression to be conserved. On the other hand, $c$ is in the 
same supermultiplet as the remaining auxiliary fields, whose boundaries
change by phases\ref{13} $\E^{i\pi\over n}$. The above association with the
supersymmetry current in order to achieve conservation would imply such
phases in the asymptotic values of the fields and charges. In fact, the low 
energy  effective equation of motion for $c$ implies a term of the form 
${i\over 4\pi}\phi_5\gamma_5 c$, leading to contributions of the type
$\overline\psi\gamma_5\psi\gamma_5$ and $\overline\psi\gamma_\mu\psi\gamma^\mu$
after use of the Ward identity.  For such boundary conditions, the
contribution from a supersymmetry current to the Wess Zumino consistency
condition, seems to be unavoidable. These issues are currently under 
investigation.

Acknowledgements: the authors would like to thank L. Alvarez-Gaum\'e, D. 
Freedman,  O. Piguet, K. Sibold, and especially R. Stora for discussions.

\vskip 2cm
\centerline {\bf References}
\vskip .5cm
\refer[[1]/D. Gepner, Commun. Math. Phys. {\bf 141} (1991)381; Nucl. Phys. 
{\bf B296} (1987)380]

\refer[[2]/C. Vafa, Mod. Phys. Lett. {\bf A4} (1989)1169]

\refer[[3]/P. Fendley, S.D. Mathur, C. Vafa and N. Warner, Phys. Lett. 
{\bf 283B} (1990)257]

\refer[[4]/C. Vafa and N. Warner, Phys. lett. {\bf B218} (1989)51]

\refer[/E. Martinec, Phys. Lett. {\bf 217B} (1989)431] 

\refer[[5]/V.I. Arnold, S.M. Guseian-Zadeh and A.N. Varchenko, {\it
Singularities of Differentiable Maps}, Birkhauser, 1985]

\refer[[6]/P. Fendley and K. Intriligator, Nucl. Phys. {\bf B372} (1992)533]

\refer[[7]/W. Lerche, C. Vafa and N. Warner, Nucl. Phys. {\bf B324} (1989)427]

\refer[[8]/Y. Kazama, H. Suzuki, Mod. Phys. Lett. A4 (1989)235;
Phys. Lett. B216(1989)112; Nucl. Phys. B321(1989)232.]

\refer[[9]/R. K\"oberle and V. Kurak, Phys. Lett. {\bf 191B} (1987)295]

\refer[[10]/M.C.B. Abdalla and A. Lima Santos, Acta Phys. Pol. {\bf B15}
(1984)813]

\refer[/M. Karowski, V. Kurak and B. Schroer, Phys. Lett. {\bf 81B} (1979)200]

\refer[[11]/E. Abdalla, M.C.B. Abdalla, N. Alves, C.E.I. Carneiro, Phys.
Rev. {\bf D41} (1990)571 ]

\refer[[12]/ O. Piguet,  and K. Sibold, Nucl. Phys. B253(1985) 269; P.L.
White, Class. Q. Gravity 9(1992) 413,1663]

\refer[[13]/A. D'Adda, P. diVecchia and M. L\"uscher, Nucl. Phys. {\bf B152}
(1979)125]

\refer[/E. Witten, Nucl. Phys. {\bf B149} (1979)285] 

\refer[[14]/R. K\"oberle and V. Kurak, Phys. Rev. {\bf D36} (1987)627]

\refer[[15]/E. Witten and D. Olive, Phys. Lett. {\bf 78B} (1978)97]

\refer[[16]/N.N. Bogoliubov and  O.S. Parasiuk, Acta Math. {\bf 97} (1957)227]

\refer[/K. Hepp, Commun. Math. Phys. {\bf 2} (1966)301]

\refer[/W. Zimmermann, {\it Lectures on Elementary Particles and Quantum Field
Theory}, Vol 1, 1970, Brandeis Univ. Summer Institute.]

\refer[[17]/J. Lowenstein, BPHZ Renormalization; Lectures given at Int. School
of Mathematical Physics, Erice, Sicily, August 17-31, 1975,
Erice Math. Phys. 1975:95, (QCD161:I651:1975). ]

\refer[[18]/E. Abdalla, M.C.B. Abdalla and K. Rothe, {\it Non Perturbative
Methods in 2D Quantum Field Theory}, World Scientific Publishing, 1991]

\refer[[19]/P. Gaigg, M. Schweda, O. Piguet and K. Sibold, Fortschritte der 
Physik 32(1984) 623]

\refer[[20]/E. Abdalla and M.C.B. Abdalla, work in progress]

\refer[[21]/ J.A. Dixon, Talk given at the Harc conference on recent advances in
the superworld, hepth/9308088]

\end